\documentclass{article} 
\usepackage{iclr2024_conference,times}

\usepackage{amsmath,amsfonts,bm}

\def\eqref#1{equation~\ref{#1}}

\def\1{\bm{1}}

\DeclareMathAlphabet{\mathsfit}{\encodingdefault}{\sfdefault}{m}{sl}
\SetMathAlphabet{\mathsfit}{bold}{\encodingdefault}{\sfdefault}{bx}{n}

\usepackage{hyperref}
\usepackage{url}
\usepackage{fancyhdr}
\usepackage{tcolorbox}
\usepackage{colortbl}
\usepackage{xcolor}
\usepackage{pifont}        
\usepackage[T1]{fontenc}   

\title{Large Language Model Evaluation Via Multi AI Agents: Preliminary results}

\author{Zeeshan Rasheed\\
Department of Computer Science\\
Cranberry-Lemon University\\
Pittsburgh, PA 15213, USA \\
\texttt{\{hippo,brain,jen\}@cs.cranberry-lemon.edu} \\
\And
Ji Q. Ren \& Yevgeny LeNet \\
Department of Computational Neuroscience \\
University of the Witwatersrand \\
Joburg, South Africa \\
\texttt{\{robot,net\}@wits.ac.za} \\
\AND
Coauthor \\
Affiliation \\
Address \\
\texttt{email}
}

\begin{document}

\maketitle

\begin{abstract}
The increasing significance of Large Language Models (LLMs) in both the academic and industrial sectors can be attributed to their exceptional performance across a diverse range of applications. As LLMs have become integral to both research and daily operations, rigorous evaluation is crucial. This assessment is important not only for individual tasks but also for understanding their societal impact and potential risks. Despite extensive efforts to examine LLMs from various perspectives, there is a noticeable lack of multi-agent AI models specifically designed to evaluate the performance of different LLMs. To address this gap, we introduce a novel multi-agent AI model that aims to assess and compare the performance of various LLMs.
Our model consists of eight distinct AI agents, each responsible for retrieving code based on a common description from different advanced language models, including GPT-3.5, GPT-3.5 Turbo, GPT-4, GPT-4 Turbo, Google Bard, LLAMA, and Hugging Face. Our developed model utilizes the API of each language model to retrieve code for a given high-level description. Additionally, we developed a verification agent, tasked with the critical role of evaluating the code generated by its counterparts. We integrate the HumanEval benchmark into our verification agent to assess the generated code's performance, providing insights into their respective capabilities and efficiencies. Our initial results indicate that the GPT-3.5 Turbo model's performance is comparatively better than the other models. Initially, we provided ten common high-level input descriptions to our proposed model. This preliminary analysis serves as a benchmark, comparing their performances side by side. 
Our future goal is to enhance the evaluation process by incorporating the Massively Multitask Benchmark for Python (MBPP) benchmark, which is expected to further refine our assessment. Additionally, we plan to share our developed model with twenty practitioners from various backgrounds to test our model and collect their feedback for further improvement.
\end{abstract}

\section{Introduction}
\label{Introduction}
Recent advancements in the field of Large Language Models (LLMs) have brought about significant changes in academia and several industries (\cite{hou2023large}). LLMs have become crucial subjects of study. For example, LLMs such as OpenAI's GPT \cite{radford2018improving}, Google's Bard \cite{thoppilan2022lamda}, Bing Chat Enterprise \cite{rudolph2023war}, LLama \cite{touvron2023llama}, and Hugging Face \cite{wolf2019huggingface} demonstrate exceptional skills in understanding, interpreting, and generating text that closely resembles human communication, proving to be invaluable assets in various domains. The importance of LLMs extends beyond their technical capabilities to their exceptional proficiency in processing and analyzing vast amounts of data effectively \cite{rasheed2024can}. This groundbreaking technology is carving new paths in both research and practical applications, particularly in the domain of Software Engineering (SE) \cite{wang2023software}.

LLMs have recently demonstrated remarkable abilities in processing and interpreting programming languages, a key aspect of source code modeling \cite{alon2020structural}, \cite{hellendoorn2017deep}, \cite{karampatsis2020big}, \cite{rasheed2024codepori}.
These models, driven by sophisticated algorithms and expansive datasets, have demonstrated an unprecedented ability to comprehend and generate code, offering substantial potential for SE applications \cite{xu2022systematic}, \cite{rasheed2023autonomous}. With the proliferation of various LLMs by leading AI research institutions and corporations, such as OpenAI's GPT series, Google's Bard, DeepMind's AlphaCode, and others, It becomes crucial to develop a strong evaluation framework to assess their performance and effectiveness in code generation tasks. 

This paper proposes a novel multi-agent AI model designed for a crucial task. The model utilizes the strengths of eight distinct AI agents, each interacting with a different LLM to retrieve programming code based on the same input. This approach aims to capitalize on the unique characteristics and methodologies of each LLM, offering a comprehensive view of the capabilities in automated code generation. We use the API key of each model to generate code. The agents work with various LLMs including GPT-4, GPT-4 Turbo, GPT-3.5, GPT-3.5 Turbo, Google Bard, LLAMA, and Hugging Face. Our proposed AI agent model facilitates code generation  and conducts initial assessments to guarantee the quality and pertinence of the results. This paper shares early insight into how the model works in practice. 

The crux of our model lies in the eighth agent, the verification agent, whose primary objective is to perform a meticulous comparison of the code produced by its seven counterparts. This agent employs HumanEval benchmark \cite{chen2021evaluating} to assesses various attributes of the generated code, such as syntactic correctness, adherence to the prompt, computational efficiency, and code accuracy. We utilized pass@k metric into our proposed AI model, enabling a direct evaluation of the code generated by various models. Pass@k metric is a method in HumanEval benchmark and a key component of the HumanEval benchmark, serves as a standard measure specifically tailored for appraising the proficiency of code generation models in creating functional and accurate code \cite{chen2022codet}. This benchmark is instrumental in providing a comprehensive and objective assessment of model performance, ensuring the reliability and effectiveness of our AI model in code generation tasks.

Our initial result demonstrates that the GPT-3.5 Turbo model exhibits superior performance in generating accurate code compared to the other models discussed earlier. We tested this by providing ten common, high-level descriptions as inputs, with each model generating code based on these descriptions. Remarkably, GPT-3.5 Turbo achieved a 70\% accuracy rate in code generation, outperforming its counterparts. Following closely, GPT-4 Turbo emerged as the second most effective model, yielding satisfactory results in 6 out of 10 cases. This study significantly contributes to the ongoing discourse on the practical applications of LLMs and aims to guide stakeholders in making informed, data-driven decisions when integrating these models into their development workflows.

Our future endeavors are focused on augmenting the evaluation process. We aim to incorporate the MBPP benchmark \cite{austin2021program} into our multi AI agent model, which we anticipate will significantly enhance the precision of our evaluations. To this end, we will develop a specialized AI agent that integrates the MBPP benchmark for a more thorough examination of the generated code. Furthermore, we intend to expand our input description set from the current 10 to 50, thereby enabling a more robust and comprehensive analysis of our proposed model. In addition to these technical enhancements, we are committed to engaging with the wider community by sharing our model with twenty practitioners from diverse backgrounds. This collaborative approach will not only provide us with valuable real-world testing but also facilitate the collection of insightful feedback, which is essential for the continuous improvement of our model. This final aspect of our plan underscores our commitment to both technical excellence and practical relevance in the evolving field of language model evaluation.

Our contribution can be summarized as follow:

\begin{itemize}
    \item Our study introduces an innovative model for code generation, offering a unique approach to the evaluation of various LLMs such as GPT-3.5, GPT-3.5 Turbo, GPT-4, GPT-4 Turbo, Google Bard, LLama, and Hugging Face.
\end{itemize}

\begin{itemize}
    \item We integrated the HumanEval benchmark into our verification agent and provided a detailed comparison of each model's code generation abilities, highlighting their strengths and weaknesses. Empirical tests with 10 common project descriptions show GPT-3.5 Turbo leading in accuracy, outperforming others in 7 out of 10 cases.
\end{itemize}

\begin{itemize}
    \item Our future goal is to enhance the evaluation process by integrating the MBPP benchmark into our specialized AI agent for comprehensive code examination, and increase input descriptions from 10 to 50 for more detailed analysis. Additionally, we will share the model with twenty diverse practitioners for real-world testing and feedback, underlining our commitment to technical excellence and practical application in language model evaluation
\end{itemize}

The rest of the paper is organized as follows: Section \ref{Related Work} reviews related work, and Section \ref{Methodology} describes the study's methodology. Primary results are presented in Section \ref{Results}, and details of our future plans are outlined in Section \ref{Enhance Evaluation}. The paper concludes in Section \ref{Conclusion}.

\section{Related Work}
\label{Related Work}
In this section, we briefly present the related work of the study, focusing on existing research. Section \ref{LLMs} provides an overview of studies concerning Large Language Models (LLMs). Section \ref{LLM Evaluation} examines works that have developed models for evaluating LLMs.

\subsection{Large Language Models}
\label{LLMs}

In recent years, LLMs have shown promise in various SE applications \cite{feng2023investigating}. LLMs are language models consisting of billions of parameters trained from a significant amount of data and have impressive performance in language processing tasks, including both natural languages and programming languages \cite{feng2020codebert}, \cite{guo2022unixcoder}. LLMs designed for processing and generating human-like text and code, ChatGPT, Google Bard, LLAMA, and Bing Chat Enterprise are prominent examples within this category \cite{eloundou2023gpts}. These LLM models are particularly notable for their ability to perform a wide range of language tasks, from translation and summarizing to question-answering and creative writing, without needing task-specific training.  
The core module behind many LLMs such as GPT and BERT are the self-attention module in Transformer \cite{vaswani2017attention} that serves as the fundamental building block for language modeling tasks. The introduction of transformer models, as exemplified by OpenAI's GPT series, marked a new era in code generation. These models, with their ability to process vast amounts of data and learn from a wide range of programming languages and styles, significantly outperformed their predecessors. GPT-3, in particular, demonstrated an unprecedented ability to generate human-like code, offering not only code completion but also bug fixing, code translation, and even generation of complex algorithms from natural language descriptions \cite{brown2020language}.

Over the past few years, LLMs have demonstrated remarkable efficiency in enabling automatic programming \cite{xu2022systematic}, \cite{sami2024system}. For example, OpenAI's suite of models has emerged as a notable force, exhibiting a profound ability in both understanding and generating code with high accuracy and efficiency \cite{radford2018improving}. Concurrently, Google's Bard and Bing Enterprises have introduced their own iterations of LLMs, tailored to enhance programming automation with distinct features and methodologies \cite{rudolph2023war}. Equally important in this space are the LLaMA and LLaMA2 models, which have carved out a niche in code generation, noted for their robustness and adaptability \cite{touvron2023llama}. Hugging Face's contributions are equally significant, offering a diverse array of models that specialize in code generation \cite{wolf2019huggingface}. These models stand out for their user-friendly interfaces and community-driven approach, making them accessible to a wider range of developers and researchers \cite{chang2023survey}. Each of these LLMs has propelled the field of automated code generation forward, presenting a variety of tools and platforms that cater to different aspects of programming and software development. This proliferation of models not only highlights the rapid advancements in AI-assisted programming but also underscores the potential for further innovation and exploration within this domain \cite{guo2023evaluating}. The comparative analysis of these models reveals a landscape rich with possibilities, offering numerous avenues for research and application in the broader context of artificial intelligence and machine learning. 

\subsection{Large Language Models Evaluations}
\label{LLM Evaluation}
It's important to assess how well LLMs perform in creating code. This helps make sure these sophisticated AI models are useful and dependable for actual software development tasks. Checking their performance is essential to confirm that the code they generate is accurate, efficient, and secure, matching the strict quality demands of the software field. Additionally, regular evaluations help in identifying areas for improvement, ensuring continuous advancement in AI-driven coding. This process not only boosts the confidence of developers in using LLMs but also contributes to the overall growth and innovation in the technology industry. Several researchers, such as \cite{xu2022systematic}, \cite{chang2023survey}, and \cite{guo2023evaluating}, have conducted surveys of existing LLM models evaluation for code across various programming languages. The primary goal of these surveys is to shed more light on the landscape of code modeling design decisions by comparing and contrasting LLM models.

The development of metrics for evaluating code generated by LLMs has seen significant advancements in recent years, particularly with the introduction of models like CodeBERT \cite{feng2020codebert}, BertScore \cite{zhang2019bertscore}, Codex \cite{chen2021evaluating}, PolyCoder \cite{xu2022systematic}, and the development of new evaluation metrics like ICE-Score \cite{zhuo2023large}. CodeBERT developed by Microsoft, which stands out as a significant milestone in code evaluation. As a bimodal model trained on both natural language and programming languages, it offers a nuanced approach to assess the semantic correctness of code snippets \cite{mashhadi2021applying}. CodeBERT's ability to understand the context and functionality of the code has made it a valuable tool for evaluating the quality of LLM-generated code. After one year. Following CodeBERT, OpenAI introduced Codex, an AI model based on GPT-3 but fine-tuned for programming tasks. Codex marked a significant step forward in code generation, capable of understanding and generating complex code sequences. Its performance necessitated the development of advanced metrics that could effectively evaluate both the syntactic and logical aspects of the code it generated \cite{chen2021evaluating}, \cite{xu2022systematic} introduce PolyCoder, an autoregressive language model specifically designed for coding, further advanced the field. PolyCoder's specialization in code generation emphasized the need for robust metrics capable of assessing intricate code structures and the model's comprehension of programming paradigms. According to \cite{xu2022systematic}, PolyCoder outperformed codex.  Furthermore, using human-written test suites to evaluate the functional correctness of LLM-generated code can be challenging, especially in domains where resources are scarce. To address these challenges and limitations, a new metric, ICE-Score, was proposed. ICE-Score revolutionizes code evaluation by instructing LLMs for code assessments, offering a novel approach to overcome the obstacles of traditional metrics \cite{zhuo2023large}. It represents a significant step forward in accurately assessing the practical utility and reliability of code produced by advanced AI models.

However, when it comes to multi-AI agent models, there is a notable gap in the development of multi-agent systems that can directly retrieve code using an API key and then evaluate the code accuracy of various LLMs. Addressing this gap, we introduce a novel multi-agent AI model designed to assess and compare the performance of different LLMs. Our work is pioneering in integrating the pass@k metric into an AI agent for a comparative analysis of code generated by these models. This approach not only enhances the precision of our evaluation but also provides a more nuanced understanding of each model's capabilities. Looking towards the future, we aim to further improve our evaluation process by incorporating the MBPP benchmark. We believe that this addition will significantly refine our assessment methods. Furthermore, to ensure the practical applicability and robustness of our model, we plan to collaborate with twenty practitioners from diverse backgrounds. By engaging these practitioners in testing our model and collecting their feedback, we seek to gain valuable insights that will guide further improvements.

\section{Research Methodology for Preliminary Analysis}
\label{Methodology}
This research aims to evaluate the performance of various language models through the use of AI agents. For this purpose, we initially developed a multi-agent model. We utilized an API key for each LLMs to obtain outputs. By providing a common description, we were able to receive code generated by different LLMs as outputs. Continuing with this approach, the methodology of our research is structured into distinct phases. Each phase is intentionally designed to rigorously test and analyze the capabilities of the LLMs in generating code responses to a common description, using our multi-agent model system.
We formulate the following research questions (RQs):

\begin{tcolorbox}[colback=green!2!white,colframe=black!75!black]
\textit{\textbf{RQ1.} How does a multi-agent AI system, with each agent utilizing a different advanced language model, compare in terms of efficiency and accuracy in generating code from a common description?}
\end{tcolorbox}
\textbf{Motivation}: The motivation behind this research question stems from the growing importance of AI in software development and the need to understand the efficacy of different AI approaches. By exploring how a multi-agent AI system, where each agent employs a distinct advanced language model, performs in generating code from a common description, we aim to shed light on the efficiency and accuracy of these models. This comparison is crucial in an era where the speed and reliability of automated coding are becoming increasingly vital. It also offers insights into the strengths and weaknesses of various models, guiding developers and researchers in selecting the most appropriate AI tools for their specific needs. This research not only contributes to the field of AI in coding but also has the potential to revolutionize the way we approach software development and problem-solving in the digital age.

\begin{tcolorbox}[colback=green!2!white,colframe=black!75!black]
\textit{\textbf{RQ2.} What criteria and methods does the verification agent employ to evaluate the performance of various language models in code generation tasks?}
\end{tcolorbox}
\textbf{Motivation}: The purpose of this research is to understand how a verification agent checks the work of different language models when they are used to write code. We want to know what methods and rules the agent uses to see if the code is good or not. This is important because as we use AI more for writing code, we need to make sure that the code is correct and works well. By learning how the agent checks the code, we can make better AI models for coding. This will help people who make software to use AI in a way that is safe and effective.


\subsection{Experimental Setup}
In this section, we detail our use of API keys from various language models for code generation. Our primary objective is to assess the accuracy and efficiency of different LLMs. Here, we explain the process of how these models generate code. In a subsequent section, we will discuss the methodology we adopted to evaluate the performance of the generated code. This approach allows for a comprehensive analysis of the capabilities of different LLMs in coding tasks.

\begin{figure*}[t]
    \centering
    \includegraphics[width=1.0\textwidth]{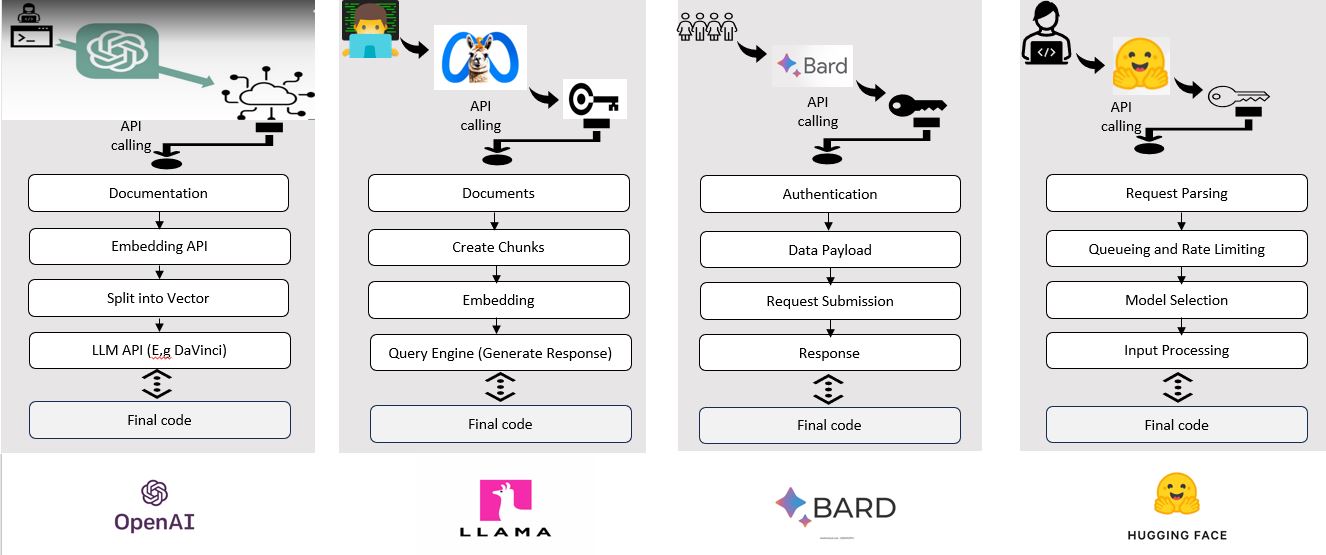}
    \caption{An overview of API request and response processes across different LLMs for Code generation}
    \label{QDA}
\end{figure*}

In this paper, we introduce a novel multi-agent AI model consisting of eight distinct AI agents. Each agent is uniquely configured to interact with a specific advanced language model, including GPT-4, GPT-4 Turbo, GPT-3.5, GPT-3.5 Turbo, Google Bard, Hugging Face, and LLAMA. The primary function of these agents is to generate code based on a common description provided to all. This ensures uniformity in the task and allows for a fair comparison of the output from each language model. The agents operate simultaneously, retrieving code based on the same set of instructions. The collected data from each agent is then prepared for subsequent evaluation. Below, we discuss the entire process of how we utilized each LLM model to generate and evaluate code from given high-level descriptions. 
Figure \ref{QDA} illustrates the entire process of how the API key is used to handle requests and generate the final code.

\subsubsection{OpenAI series}
We developed an AI agent model that integrates with multiple versions of the OpenAI GPT models, including GPT-4, GPT-4 Turbo, GPT-3.5, and GPT-3.5 Turbo. Each AI agent in our model was configured to interact with a specific version of the OpenAI GPT model. We initiated the process by providing customized prompts to each agent, embedding a particular project description into these prompts to maintain consistency across different models. The AI agents, each linked to a distinct GPT version, were tasked with processing these prompts and generating responses.

Our model efficiently managed the communication with the OpenAI API for each agent, ensuring that the correct version of the GPT model was utilized for each request. Figure \ref{QDA} demonstrates how the API key receives requests and generates responses. The API key provided secure access to these models, enabling our agents to send and receive data. The responses from the various GPT versions were then collected and processed to maintain a coherent and continuous dialogue flow.
By structuring the interaction in several rounds and alternating between the different GPT models, our model was able to capture a diverse range of AI-generated responses. This setup was pivotal in assessing the comparative performance and output variations among the different models. The final outcome of this methodology was a comprehensive collection of responses from each GPT model, offering a unique opportunity to analyze and understand the nuances in AI-generated content based on the model version and initial input conditions.

\subsubsection{Google Bard}
Google Bard, a sophisticated language model created by Google, played a crucial role in our research methodology. We employed the Google Bard API to produce descriptive content. Our method involved using the Bard API alongside Bard Cookies for interaction with Google Bard's AI system. This process necessitated authentication via specific cookies, supplied in the form of a dictionary to the Bard Cookies class. After successful authentication, the system could process and respond to text-based queries. For example, we input a high-level description into the Bard system, which then generated code relevant to the input. The responses from Bard offered insightful reflections on the AI's ability to comprehend and create intricate, contextually relevant textual content, showcasing its effectiveness in automated content creation and AI-centric research methodologies.

\subsubsection{LLAMA}

LLAMA, another prominent language model in our study, was accessed through its designated API key. This key enabled our algorithm to send high-level descriptions to LLAMA, which then employed its language understanding capabilities to produce code. We incorporated the use of the replicate Python library to interact with advanced AI models for code generation. Specifically, we utilized this library to communicate with the Meta LLAMA-2-70b model. The process involved sending a high-level prompt to the AI model and model processed the request and generated a corresponding output. This output was then iteratively printed, providing us with AI-generated Python code. This approach demonstrated the model's capability to understand and generate complex code based on simple text prompts, showcasing its potential in automating the code generation process and aiding in software development research.

\subsubsection{Hugging Face}
In our research, we also leveraged Hugging Face's extensive suite of language models through its API, focusing specifically on the CodeBERT model. Our method involved providing a high-level description to the model. Utilizing its advanced natural language processing capabilities, Hugging Face's model generated relevant code. This interaction was made possible through the API key, which served as a conduit, granting our algorithm access to various models under the Hugging Face platform. Each model contributed distinct strengths to the code generation process, demonstrating the versatility and efficacy of Hugging Face's AI solutions in our research.

\subsection{Input of Code Evaluation}
In this research, we focused on the development of an AI agent model capable of robustly interpreting natural language prompts and generating corresponding code based on the provided descriptions. Our methodology involved utilizing common natural language descriptions as inputs, which were then processed by various state-of-the-art LLMs to generate code outputs.

To effectively evaluate the performance of these LLMs in code generation, we curated a diverse set of 10 descriptive prompts, each originating from different project backgrounds. This diversity in input descriptions was deliberately chosen to encompass a wide range of scenarios and complexities that developers might encounter in real-world applications. The specifics of these input descriptions are presented below. 

The box below displays the input prompts used for evaluating various LLMs, thereby ensuring a comprehensive assessment of the LLMs' code generation capabilities. By encompassing a wide array of subjects and complexities in our input descriptions, our approach aims to rigorously test the adaptability and accuracy of the LLMs in converting natural language instructions into functional code across different contexts. This methodological approach provides a robust framework for assessing the efficacy of LLMs in understanding and translating human language into executable code.

\begin{tcolorbox}[colback=gray!2!white,colframe=black!75!black]
\textit{\textbf{01.} A simple Python program for creating and using flashcards for study, allowing users to test their knowledge on various subjects}\\
\textit{\textbf{02.} An interactive Python game that tests users on various academic subjects and provides instant feedback}\\
\textit{\textbf{03.} A short, interactive, text-based adventure game written in Python, where players make choices that influence the outcome of the story.}\\
\textit{\textbf{04.} A Python-based Pomodoro timer that helps users apply the Pomodoro technique (25 minutes of work followed by a 5-minute break) to boost productivity.}\\
\textbf{.}\\
\textbf{.}\\
\textbf{.}\\
\textbf{.}\\
\textit{\textbf{10.} A Python-based basic arithmetic calculator capable of performing operations like addition, subtraction, multiplication, and division}
\end{tcolorbox}


\subsection{Methodology for Code Evaluation}
The evaluation process involves analyzing various aspects of the code generated by each language model. Key metrics include the accuracy of the code in relation to the given description, efficiency in terms of execution and resource usage, and the innovation demonstrated in the solutions. This study employs a structured approach to assess the code generation capabilities of various advanced LLMs, including GPT-4, GPT-4 Turbo, GPT-3.5, GPT-3.5 Turbo, LLAMA, Hugging Face, and Google BARD. We developed a specialized AI agent with the primary objective of evaluating the accuracy and reliability of code produced by these models. The cornerstone of our evaluation methodology is the integration of the HumanEval benchmark within our AI agent.

This model utilized the HumanEval benchmark, a widely recognized standard for assessing code quality and effectiveness. For evaluation, we employed the pass@k metric, a rigorous method to determine the viability of generated code. Our model receives a description as input, and in response, it obtains code from various LLMs. We provided ten distinct descriptions, each generating a set of potential solutions. Using the pass@k metric, we evaluated these solutions to determine their correctness. Mathematically, pass@k can be expressed as the probability that at least one of the top k solutions generated by the model is correct.

\[
\text{pass}@k := \mathbb{E}_{\text{Problems}} \left[ 1 - \frac{\binom{n-c}{k}}{\binom{n}{k}} \right]
\]

If one solution out of the k attempts is found correct, we mark the instance as successful and proceed to the next description. In this way, we systematically assess the performance of LLMs across multiple coding tasks, ensuring a comprehensive analysis of their capabilities in code generation. The methodology's implementation involves iterating over the set of descriptions, generating code for each, and applying the pass@k metric to assess each set of solutions. This approach allows for an empirical evaluation of LLMs in terms of both accuracy and efficiency in coding tasks.

\section{Preliminary Results}
\label{Results}
In this section, we present the initial results of the proposed multi agents model for evaluation of various LLM-based model. The primarily results are shown in Table \ref{tab:LLM evaluation}. So far our findings indicate that the GPT 3.5 Turbo model perform relatively better than others LLM based model for generating accurate code by given high level description. Below, we present the results of our proposed model in Section \ref{LLM based model RQ1}, specifically reporting the outcomes of RQ1. Additionally, we also provide the detail criteria and methods that we employed in verification agent to evaluate the performance of various language models in Section \ref{Initial Evaluation Framwork}.
\subsection{Proposed Multi Agent Model (RQ1)}
\label{LLM based model RQ1}
In this section of our research, we conducted an evaluation of code performance across various language models. This assessment was carried out by developing a sophisticated algorithm that utilized the API keys of the respective language models. The core objective was to methodically analyze and compare the output quality and efficiency of each model when tasked with generating code from high-level descriptions. In the pursuit of understanding the capabilities of various LLMs in the context of code generation, we conducted a comprehensive evaluation using the HumanEval benchmark to gauge the accuracy of the code produced by each model. Table \ref{tab:LLM evaluation} presents a synthesized view of this evaluation, showcasing the performance of different LLMs across identical input descriptions.

Upon early testing and analysis, our results indicated that GPT-3.5 Turbo exhibited superior performance compared to other language models like GPT-4, GPT-4 Turbo, Google Bard, Hugging Face, and LLAMA. Specifically, GPT 3.5 Turbo provided accurate results for seven out of the ten inputs. This achievement marks a significant margin of excellence compared to its counterparts. In contrast, other LLM models in the study produced fewer accurate outputs, as can be observed from the comparative data in the Table \ref{tab:LLM evaluation}. 

Table \ref{tab:LLM evaluation} enumerates seven different LLMs, each produced by renowned organizations with varying parameter sizes—from 355M by Hugging Face to 1.96 trillion by GPT-4 Turbo. Each model was provided with the same set of ten input descriptions to process and generate code. The primary focus of the evaluation was on the accuracy of the results, which refers to the functional correctness of the generated code as per the given description. Additionally, a quality rating, depicted with stars, provides a subjective assessment of the code based on criteria such as readability, efficiency, and adherence to best practices.

The data shows a varied performance across models. GPT-3.5 Turbo, with 154 billion parameters, leads with the highest number of accurate results, successfully generating correct code for seven out of ten descriptions. This is followed closely by GPT-4 Turbo, which each returned six accurate pieces of code. Notably, despite Google Bard’s significantly larger parameter count, it did not outperform GPT-3.5 Turbo, indicating that larger model size does not necessarily equate to better performance in specific tasks such as code generation.

The results underscore the efficacy of GPT-3.5 Turbo in this domain, with a strong blend of accuracy and high quality, as reflected by its four-star rating. The other models, while demonstrating varying levels of proficiency, highlight the competitive landscape of LLMs and their potential for software engineering applications.

This analysis provides initial insights into the current state of LLMs in code generation tasks, underscoring the importance of specialized benchmarks like HumanEval in evaluating AI agents. It also sets the stage for further discussions on model optimization, the impact of parameter scale, and the potential for LLMs to revolutionize the field of automated code generation.

\subsection{Initial Evaluation Framework (RQ2)}
\label{Initial Evaluation Framwork}
In the Initial Evaluation Framework section, our focus shifted towards the validation and verification of the code generated by these models. In our study, we deployed an AI multi-agent model to evaluate the code generation capabilities of various LLMs, including GPT-4, GPT-4 Turbo, GPT-3.5, GPT-3.5 Turbo, Google Bard, LLaMA, and Hugging Face. The performance assessment of these models was based on their ability to generate accurate and functional code in response to a set of predefined descriptions. 

Our evaluation strategy involved presenting each LLM with 10 unique code generation tasks. The performance of each model was assessed using the HumanEval benchmark, which is a widely recognized method for evaluating code synthesis. In this project, we developed multi AI agent model to generate and evaluate the generated code. The task of one agent was to employ the HumanEval benchmark and autonomously test the code generation capabilities of AI models Each task in HumanEval requires the AI to write a function that satisfies a given specification, and the models' responses are evaluated based on their correctness and efficiency.

One of the key metrics used in our assessment was the pass@k metric, a standard measure for evaluating code generation models. The pass@k metric evaluates whether a correct solution is found within k attempts, where k is a predefined number of generated code samples. This metric is particularly useful in scenarios where multiple solutions are possible, as it accounts for the probabilistic nature of code generation by LLMs.
In our experiment, we utilized a pass@1 approach, meaning that we evaluated the models based on their first attempt at generating a solution. This approach was chosen to reflect a more realistic scenario where a developer would use the model's first output. 

The results were revealing: GPT-3.5 Turbo outperformed other models in our tests, delivering correct and efficient code solutions in 7 out of 10 tasks. This indicates a 70\% success rate on the first attempt, which is a significant achievement considering the complexity of the tasks involved. In comparison, other models like GPT-4 and GPT-4 Turbo showed varying levels of performance, but none matched the consistency of GPT-3.5 Turbo.

This evaluation highlights the effectiveness of GPT-3.5 Turbo in code generation tasks, particularly in its ability to understand and translate complex task descriptions into functional code. The use of HumanEval benchmark and the `pass@k` metric provided a robust framework for this assessment, allowing for a fair and comprehensive comparison across different LLMs.

\begin{table*}[ht!]
\centering
\caption{Workflow of the proposed model: Application across diverse datasets and output formats}
\label{tab:LLM evaluation}
\begin{tabular}{|
>{\columncolor[HTML]{EFEFEF}}l |
>{\columncolor[HTML]{EFEFEF}}l |
>{\columncolor[HTML]{EFEFEF}}l |
>{\columncolor[HTML]{EFEFEF}}l |
>{\columncolor[HTML]{EFEFEF}}l |
>{\columncolor[HTML]{EFEFEF}}l |}
\hline
\textbf{LLM Model} & \textbf{Product} & \textbf{Parameter} & \textbf{I/P Description} & \textbf{Accurate Result} & \textbf{Quality} \\ \hline
GPT-4 Turbo        & OpenAI           & 1.96 trillion       & 10                       & 6                        & \ding{72}\ding{72}\ding{72}\ding{73}\ding{73}        \\ \hline
GPT-4              & OpenAI           & 1.76 trillion      & 10                       & 5                        & \ding{72}\ding{72}\ding{72}\ding{73}\ding{73}        \\ \hline
GPT-3.5 Turbo      & OpenAI           & 154 billion         & 10                       & 7                        & \ding{72}\ding{72}\ding{72}\ding{72}\ding{73}        \\ \hline
GPT-3.5            & OpenAI           & 125 billion        & 10                       & 5                        & \ding{72}\ding{72}\ding{72}\ding{73}\ding{73}        \\ \hline
Google Bard        & Google           & 1.56 trillion      & 10                       & 4                        & \ding{72}\ding{72}\ding{73}\ding{73}\ding{73}        \\ \hline
LLama              & Meta             & 70 billion         & 10                       & 2                        & \ding{72}\ding{73}\ding{73}\ding{73}\ding{73}        \\ \hline
Hugging Face        & Hugging Face    & 355M               & 10                       & 2                        & \ding{72}\ding{73}\ding{73}\ding{73}\ding{73}        \\ \hline
\end{tabular}
\end{table*}

\section{Future Work}
\label{Enhance Evaluation}
As we work to improve how we evaluate LLMs for code generation, our current methodology employs the HumanEval benchmark within our AI agent framework. HumanEval has served as a robust starting point, offering a suite of programming problems that models must solve, thereby providing a quantitative measure of their code generation abilities. However, to broaden our evaluation's scope and depth, we are preparing to incorporate the MBPP. The MBPP benchmark, with its larger and more diverse set of programming tasks, will enable a more extensive assessment of LLMs' capabilities. 

In conjunction with the adoption of MBPP, we recognize the indispensable value of human judgment in model evaluation. To this end, we have embarked on a collaboration with twenty practitioners from a spectrum of backgrounds. These industry experts, academics, and developers will bring their unique perspectives to bear on the task of manually evaluating our proposed model's performance. Their insights will not only validate the findings from the MBPP benchmark but also provide nuanced understanding of the model's practical effectiveness and areas for improvement.

This dual approach, combining automated benchmarks with human expertise, wil offer a comprehensive evaluation framework. It will ensure that the strengths and weaknesses of LLMs in code generation are fully understood, paving the way for targeted enhancements. With this initiative, we aim to set a new standard for the evaluation of LLMs, one that balances the scalability of automated testing with the depth of human review.

\section{Conclusions}
\label{Conclusion}

In this paper, we introduced a novel multi-agent AI model designed to assess and compare the performance of various LLMs. Our system comprises eight distinct AI agents, each tasked with the retrieval of code based on a shared high-level description from several advanced language models, including GPT-3.5, GPT-3.5 Turbo, GPT-4, GPT-4 Turbo, Google Bard, LLAMA, and Hugging Face. Utilizing the APIs provided by these models, our agents can retrieve code that is then evaluated by a specialized verification agent. This verification agent employs the HumanEval benchmark to measure performance, offering valuable insights into the strengths and efficiencies of each model. Initial findings have highlighted GPT-3.5 Turbo's superior performance relative to its counterparts.

As we look to the future, our ambition is to further enhance the evaluation framework. The integration of the MBPP represents a significant next step in this process, promising to sharpen our assessment tools and provide even more granular insights. Moreover, we are committed to engaging with the broader community by involving twenty practitioners from a variety of backgrounds to interact with and evaluate our model. Their feedback will be indispensable, providing a practical edge to our theoretical findings and helping to guide the model's continuous improvement.

Our work lays the groundwork for a dynamic and robust evaluation platform that not only benchmarks the current state of LLMs but also adapts and evolves with the field. The insights gained through our research will serve as a cornerstone for future advancements in AI-driven code generation, facilitating developments that are as practical as they are innovative. Through rigorous assessment and community engagement, we aspire to drive the field forward, setting new standards for performance and applicability in the ever-growing expanse of language model technology.

\section{Acknowledgment}
We express our sincere gratitude to Business Finland for their generous support and funding of our project. Their commitment to fostering innovation and supporting research initiatives has been instrumental in the success of our work.

\bibliography{iclr2024_conference}
\bibliographystyle{iclr2024_conference}

\end{document}